\documentclass{article}
\usepackage{spconf,amsmath,graphicx}

\usepackage{amsmath}
\usepackage{subfigure}
\usepackage{booktabs}

\title{Affective Burst Detection from Speech using Kernel-fusion Dilated Convolutional Neural Networks}
\name{Berkay Köprü and Engin Erzin
}
\address{KUIS-AI Laboratory \\
  Multimedia, Vision and Graphics Group \\
  College of Engineering, Ko\c{c} University, Istanbul, Turkey\\
  {\it bkopru17,eerzin@ku.edu.tr}
  }
\begin{document}
\ninept
\maketitle
\begin{abstract}
As speech-interfaces are getting richer and widespread, speech emotion recognition promises more attractive applications. In the continuous emotion recognition (CER) problem, tracking changes across affective states is an important and desired capability. Although CER studies widely use correlation metrics in evaluations, these metrics do not always capture all the high-intensity changes in the affective domain. In this paper, we define a novel affective burst detection problem to accurately capture high-intensity changes of the affective attributes. For this problem, we formulate a two-class classification approach to isolate affective burst regions over the affective state contour. The proposed classifier is a kernel-fusion dilated convolutional neural network (KFDCNN) architecture driven by speech spectral features to segment the affective attribute contour into idle and burst sections. Experimental evaluations are performed on the RECOLA and CreativeIT datasets. The proposed KFDCNN is observed to outperform baseline feedforward neural networks on both datasets.
\end{abstract}

\section{Introduction}

Human emotional system is driven by biological stimuli from the external or internal stimuli and interacts with the cognition system in the brain \cite{Ledoux1989}. Since human behaviours are driven by the emotional system, automatic recognition of emotions has become an attractive and important research field in the last two decades.
Speech Emotion Recognition (SER) has drawn significant attention in the recent literature as speech signal is highly correlated with the emotional states, lower-dimensional and occlusion free compared to visual modalities \cite{ser_survey1,ser_survey2,speechAdvantages1,speechAdvantages2}. SER also keeps attracting attention with the surge of speech interfaces in the Internet-of-Things applications \cite{ser_survey1}.




Discrete emotion recognition (DER) is a subclass of SER, which focuses on the categorical representation of emotions such as anger and happiness. In DER studies, audio signal is widely represented with low-level descriptors (LLDs), such as pitch, energy, zero-crossing rate and spectral features \cite{sder1,sder2}. Classification task in DER has been addressed using LLD features by hidden Markov model based approaches as in \cite{sder1} and also by recurrent neural networks (RNNs) based approaches as in \cite{sder2}. Alternatively in \cite{sderEnd2End1}, raw speech is processed with multiple 1-dimensional convolutions to classify emotions. 

Alternatively, continuous emotion recognition (CER) represents emotions in a 3-dimensional continuous affect attribute space whose dimensions are Arousal, Valence and Dominance, respectively representing activeness - passiveness, positiveness - negativeness and dominance-submissi\-ve\-ness  \cite{contEmotions}. CER studies use both tailored features as in \cite{contER_CCC1,kopruCERMTL2020} or learned features as in \cite{cerEndToEnd,cerEnd2End2,scer_addition_end2end}. In \cite{contER_CCC1}, 23 LLDs from eGeMAPS \cite{egemaps} are extracted from the audio signal. Then, based on these features, a stacked long short-term memory (LSTM)-RNNs model is proposed for CER. 
In \cite{kopruCERMTL2020}, audio signal is represented by combination of the Mel-frequency Cepstral Coefficients (MFCCs), delta and acceleration of MFCCs. Then using multi-task learning Arousal, Valence, and Dominance attributes are estimated in parallel. 
Trigeorgis et al. propose a convolutional recurrent neural network where two convolutional layers are acted as a learned feature extractor on the raw audio signal \cite{cerEndToEnd}. Similarly, Tzirakis et al. adopt convolutional neural networks (CNNs) to produce audio embeddings for CER \cite{cerEnd2End2}.
In CER studies, correlation based metrics are widely used for evaluation tasks, since the trend of the predicted attribute is accepted to be more important than the actual level of the prediction. On the other hand, correlation metrics do not always grant effective capturing of all high-intensity changes in the affective domain.  

Due to the categorized nature of the DER, while inter-emotion transitions can be observed, intra-emotion transitions do not appear or are not typically available for the DER studies. On the other hand, continuous affect attributes in the CER problem provide the necessary intensity fluctuations for the inter- and intra-emotion transitions.

The detection of inter- and intra-emotion transitions, i.e., affective change detection, has a significant importance, and it is widely studied in the psychology domain under the \textit{mismatch negativity} (MMN) literature \cite{vmmn1,vmmn2,vmmn3}. The information processing capability of human beings is limited, so automatic detection of relevant stimuli is crucial to orient attention \cite{vmmn_sup}. As the emotional expressions drive the communication for possible threats in the environment, even when attention is engaged in a concurrent task, emotional information is prioritized, and automatically processed \cite{vmmn_sup2}. Hence, following changes in the affective domain can be critically important to design natural human-computer interaction applications.

Affective change detection is studied in the DER context by \cite{affectiveChangePointML1} and CER context by \cite{affectiveTrendDetection}. Affective change points are defined as the transition points between the emotions in \cite{affectiveChangePointML1}. These points are estimated using a Gaussian mixture model based architecture with and without prior emotion class information. In \cite{affectiveTrendDetection}, emotional hotspots are defined as sections deviated from the median of the affective attribute. They proposed a qualitative agreement-based assessment method to map affective attributes into low, high, neutral, and non-consensus sections. A section is labeled as a low if that section under the median, and high if it is above the median. Then, bidirectional long short-term memory (BLSTM) is operated on 88 eGeMAPS \cite{egemaps} features, which are extracted from acoustic signals, to classify the trend. However, this approach highlights flat sections that are deviated from the median, and it resembles a quantization approach more than tracking the high-intensity regions as the labeling procedure misses all the inter- and some of the intra-emotion transition regions in the affective domain. 


In this study, we address segmentation of the affect contour into affective burst and idle regions. Unlike \cite{affectiveTrendDetection}, we label sections regarding the intensity of change. Affective burst sections then estimated using spectal features of speech as input and a kernel-fusion dilated convolutional neural network (KFDCNN) as the classier. To summarize, the main contributions of this study are as follows:
\begin{itemize}
\itemsep0em 
    \item We propose a new labelling mechanism to define high intensity and idle segments of the affect contour.
    \item We formulate a novel affective burst detection problem capturing the high intensity changes, which can lead to improve the understanding of the inter- and intra-emotion transitions in the scene.
    \item We propose a novel architecture KFDCNN for affective burst detection from speech.
    \item We carry out evaluations on the RECOLA and CreativeIT datasets with classification metrics.
\end{itemize}

\begin{figure}[htbp]
\subfigure[Labeled affective burst segments for Arousal from RECOLA dataset]{
\includegraphics[width=0.95\columnwidth]{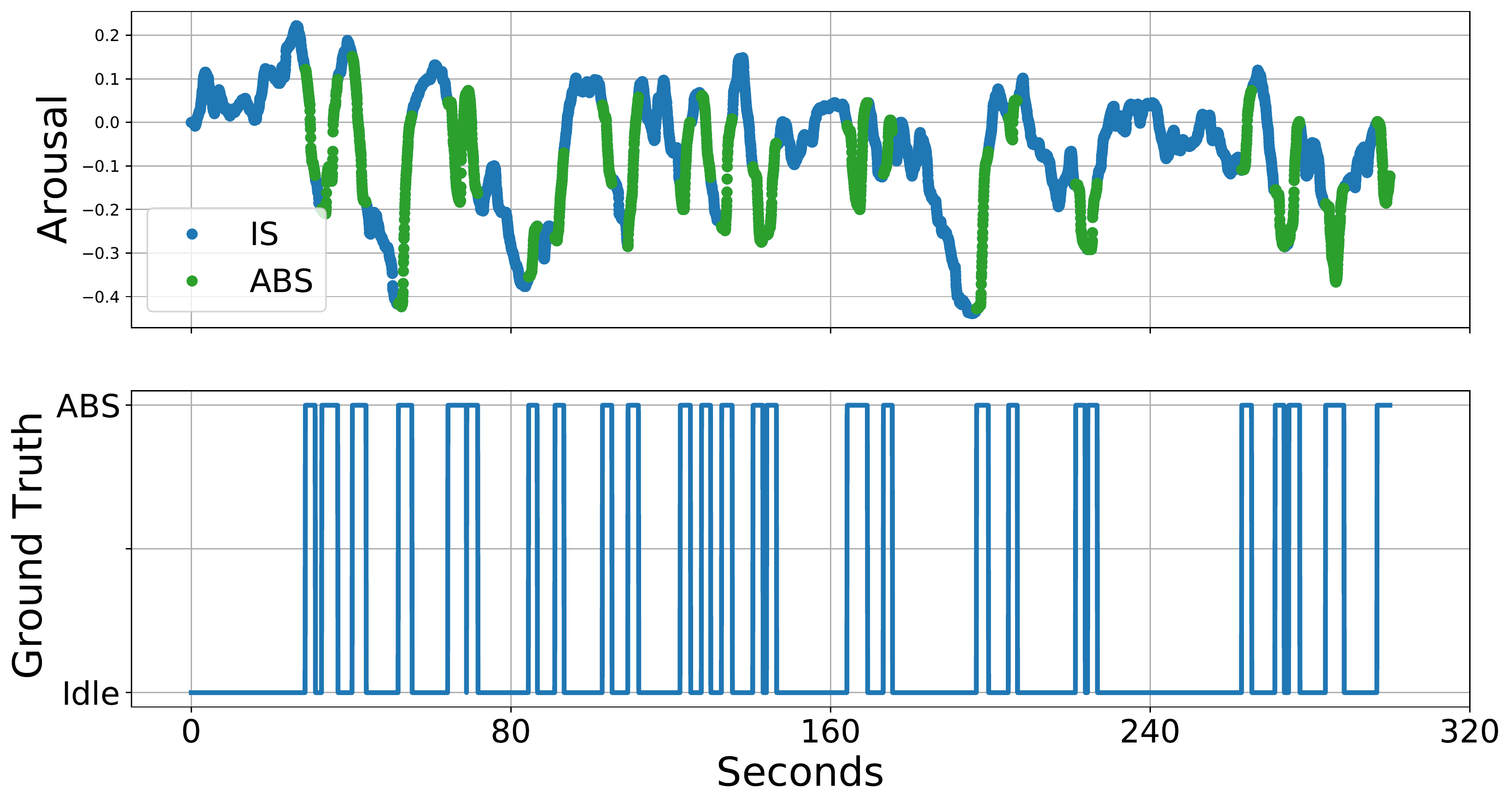}
} 
\subfigure[Labeled affective burst segments for Valence from CreativeIT dataset]{
\includegraphics[width=0.95\columnwidth]{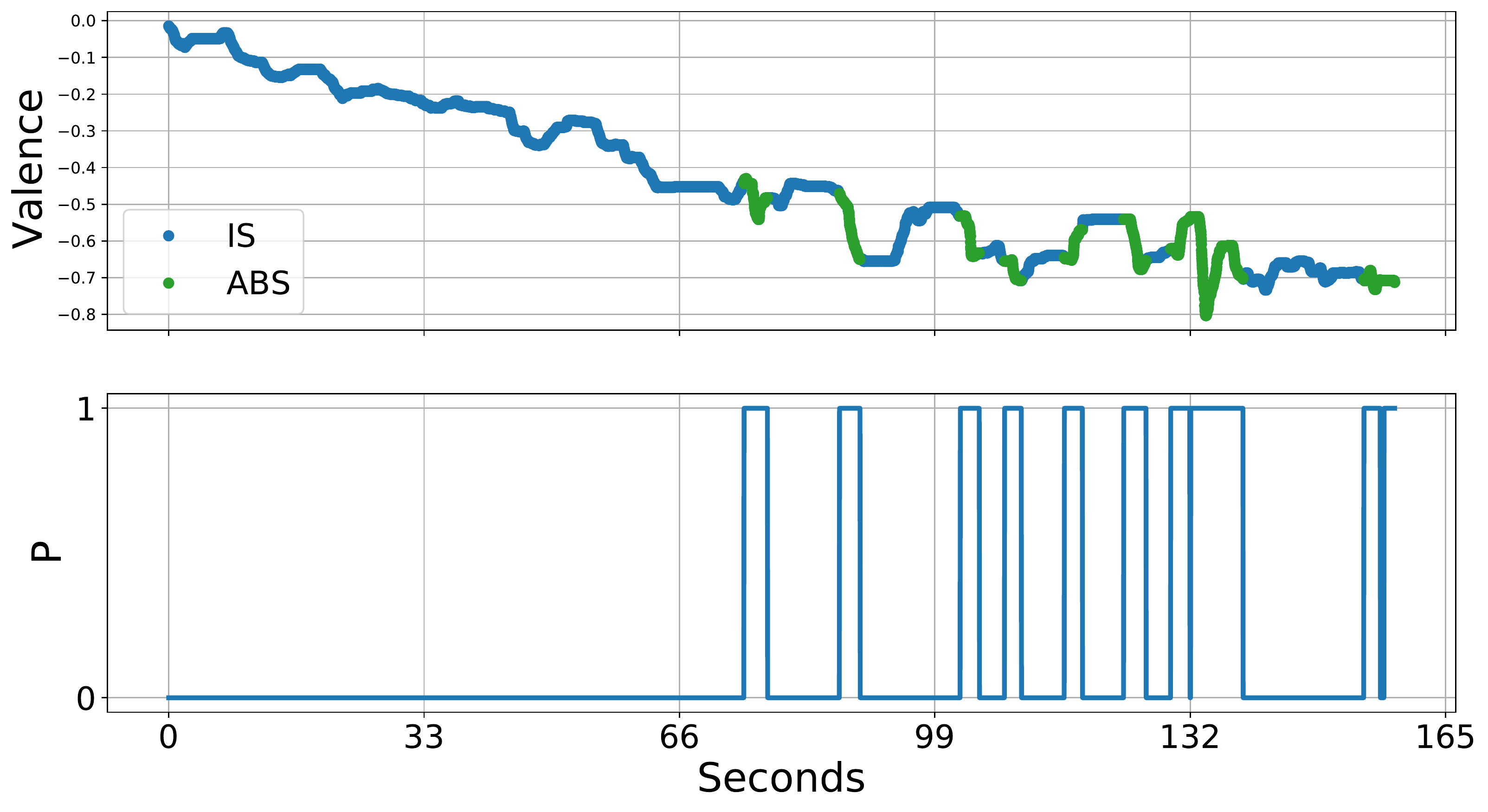}
}
\caption{Sample Arousal and Valence contours with affective burst and idle segment labelings}
\label{fig:annotation}
\end{figure}

\begin{figure*}[!ht]
\centering 
\includegraphics[scale=0.65]{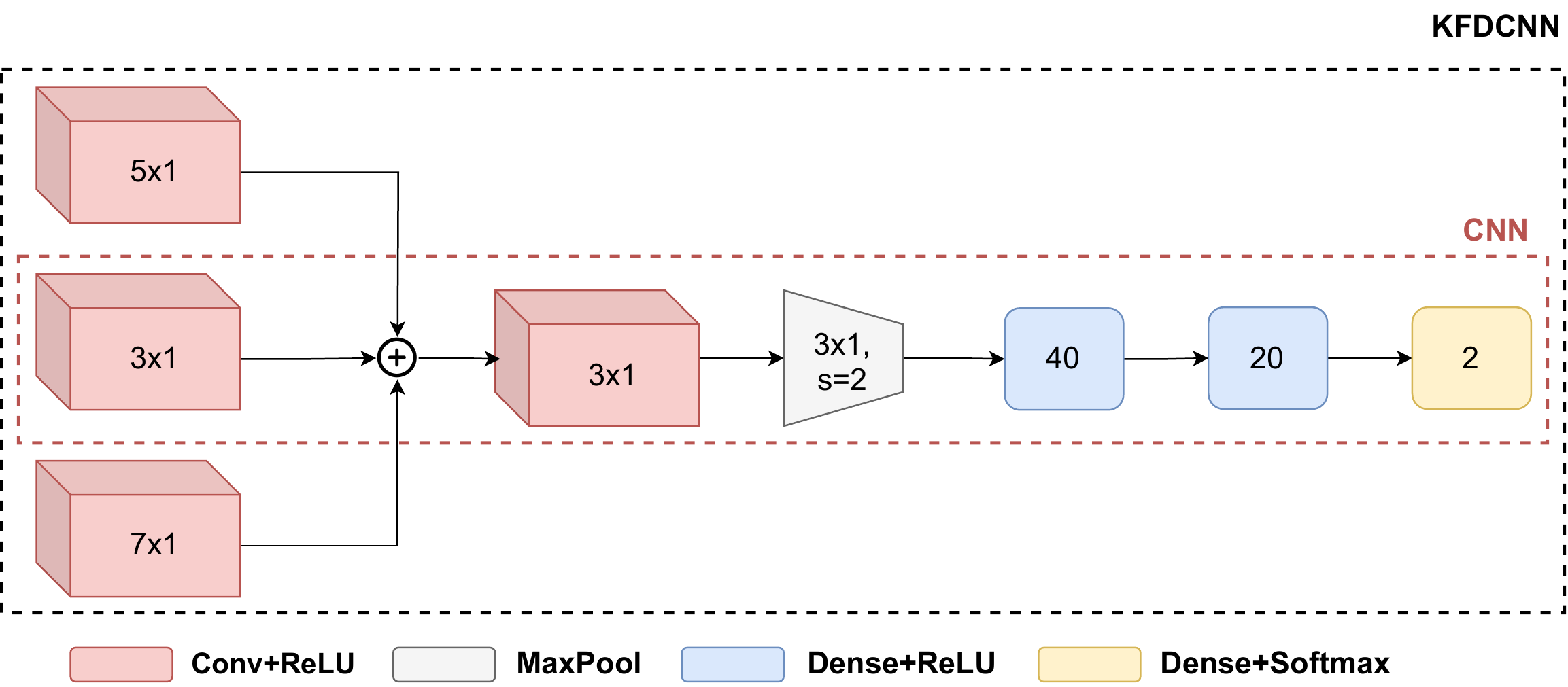}
\caption{Kernel fusion dilated convolutional neural network for affective burst detection}
\label{fig:KFDCNN}
\end{figure*}

\section{Method}

We propose a convoltional neural network (CNN) based architecture where the input is processed by parallel convolutions having different dilated kernel lengths to detect affective bursts. In this section first, affective burst detection problem is defined, then feature extraction from speech signals is described. Later, affective burst detection framework based on kernel fusion and CNNs are described.


\subsection{Affective Bursts}

Affective burst detection is a two-class classification problem where the affective contour is segmented into affective burst and idle classes. We define the affective burst segment (ABS) as the region in which the affect attribute contour is changing rapidly with high gradients. Respectively, idle segments (ISs) cover complement of the ABSs and correspond to the regions where the affect attribute contour is changing slowly with low gradients.

The ground-truth ABS annotations are generated in two steps. First affective burst points (ABPs) on the affect attribute contours are detected, then these points are extended into segments which are referred as ABSs. Note that all non-ABS regions are referred as idle segments.

Affective burst points (ABPs) are set based on the first order regression coefficients of the arousal and valence attributes as
\begin{equation}
    d_e[n] =  \frac{\sum_{l=1}^L(e[{n+l}]-e[n-l])l}{2\sum_{l=1}^{L}l^2}
\end{equation}
where $d_e[n]$ is the delta coefficient of attribute $e$ (Arousal or Valence) at sample index $n$. For simplicity, we will drop $e$. By selecting a threshold $\tau$, we can define the ABP indicator function $p$ at sample index $n$ as
\begin{equation}
  p[n] = 
 \begin{cases}                   
 0 & \text{if $ -\tau < d[n] < \tau$} \\
 1 & \text{if $ | d[n] | \geq \tau$}.
 \end{cases}
 \label{eq::annotation}
\end{equation}

Then the ground truth binary segment labels are extracted as
\begin{equation}
P[n+i] = 
 \begin{cases}                   
 1 & \text{$\forall n$ s.t. $p[n] = 1$ and $i=-\Delta,\dots,\Delta$} \\
 0 & \text{otherwise}
 \end{cases}
  \label{eq::annotation2}
\end{equation}
where an ABS of temporal size $w_s= 2\Delta+1$ is centered for each ABP on the indicator function $p[n]$. Note that the resulting size of ABSs can be longer than $w_s$ when two consecutive ABPs are closer than $w_s$. Sample affect attribute contours for arousal and valence, indicated as ABS and IS in different colors, together with $P[n]$ are depicted in Figure~\ref{fig:annotation}. 





\subsection{Feature Extraction}

In this study, we use the \textit{extended Geneva Minimalistic Acoustic Parameter Set (eGeMAPS)} representing the spectral and temporal characterization of speech signal for ABS detection \cite{egemaps}. The 88 dimensional eGeMAPS features are calculated as statistics of 25 \textit{low-level descriptors} (LLDs), such as  Mel-Frequency Cepstral Coefficients, Pitch and Loudness, using the OpenSMILE toolkit \cite{opensmile}.

The LLDs are extracted using window size of 20~ms, and the 88 dimensional eGeMAPS features are calculated over 500~ms with a hop duration of 40~ms. Hence the eGeMAPS features are extracted at 25~fps and represented at time frame $n$ as $F[n] \in \mathbf{R}^{88}$. 

\subsection{Kernel Fusion Convolutional Neural Network}

The proposed KFDCNN architecture is depicted in Figure~\ref{fig:KFDCNN}. KFDCNN is composed of both dilated parallel and typical convolutional layers, max-pooling layer, and multiple fully connected layers. Kernel fusion layer at KFDCNN includes dilated convolutional kernels having different lengths that helps to learn temporal relations in different resolutions and enriches representation for the ABS detection. 
Furthermore, with dilated kernels each layer has longer receptive fields, resulting in convolution outputs that capture long-term information which is especially important for slowly varying emotional processes. 

The proposed KFDCNN architecture processes a window of features which is represented as $I[n] = \{F[n-T], F[~n~-~T~+~s], \ldots, \\\ F[n+T-s], F[n+T]\}$ where $s$ is the dilation rate, and $I[n] \in \mathbf{R}^{\frac{2T}{s} \times 88}$. Then KFDCNN outputs $\hat{y}[n] \in \mathbf{R}^{2}$ from the input $I[n]$.

\subsection{Model Training}
The architecture is trained for ABS detection through a binary classification task. However, this task has an imbalanced nature. There are a small number of sections labeled as ABS, while a high number of sections are labeled as IS. To overcome the imbalance problem, we adopted weighted negative log-likelihood ratio loss: 
\begin{align}
    L(\mathbf{P},\mathbf{\hat{y}}, \mathbf{\theta}) = \sum_n \frac{1}{(\theta_0+\theta_1)} - \theta_{P[n]} \log(\hat{y}[n, P[n]])
\end{align}
where $\theta_i$ is the weight for class $i$, $P[n] \in \{0,1\}$ is the binary segment label at time frame $n$, $\hat{y}[n, P[n]]$ is the $P[n]$-th component of the KFDCNN output $\hat{y} \in \mathbf{R}^{B\times 2}$ at frame $n$, and $B$ is the batch size. The class weight $\theta_i$ is extracted as
\begin{align}
    \theta_i = \frac{1}{\text{frequency of $i^{th}$ class}}
\end{align}
for class index $i=0, 1$ corresponding to the IS and ABS.

\section{Experimental Evaluations}

The proposed architecture in Section~2 is evaluated on the RECOLA \cite{recola} and the CreativeIT \cite{creativeIt} datasets.  In this section, datasets and implementation details are introduced. Then evaluation metrics are described. Finally, performance of the KFDCNN is compared against a baseline feed-forward neural network (FFN) together with the CNN and the dilated CNN (DCNN).

\subsection{Datasets}

We train and evaluate the ABS detection task on the widely used multi-modal datasets RECOLA and CreativeIT. RECOLA dataset is composed of multi-modal recordings of dyadic conversations of 27 French speakers. As a part of AVEC16 challenge, the dataset is divided into uniform-sized training, development, and test sets. While annotations for the training and the development sets are available, the annotations for the test set are not public. Publicly available annotations are for the arousal and valence attributes at 25~Hz rate. 

USC CreativeIT is a multimodal database of theatrical improvisations. Each interaction on average has a length of 3.5~minutes and is captured by recordings of the body Euler angles and speech from the participants. The dataset includes references for arousal, valence and dominance. It is divided into 5 sessions which are mutually exclusive in terms of speakers. The cross-validation procedure on this dataset is held by leave-one session out to preserve speaker independence. 

\begin{table}[!htb]
\begin{center}
\caption{Statistics over the RECOLA and CreativeIT datasets after the ground-truth ABS annotations on Arousal (A) and Valence (V) contours: Number of ABSs, mean duration of ABSs, total duration of ABSs, and mean absolute delta ($|d|$) of the ABSs}
\label{tab:dsets}
    \begin{tabular}{l|c c |c c}
    \toprule[1pt]\midrule[0.3pt]
    \textbf{Stats} &\multicolumn{2}{c|}{\textbf{RECOLA}} & \multicolumn{2}{c}{\textbf{CreativeIT}}\\
    & A  & V & A  & V \\ \midrule \hline 
\# ABSs & 446 & 461 & 641 & 632 \\ \hline 
Mean ABS dur (sec) & 3.4 & 3.8 & 3.6 & 3.6\\ \hline
Total ABS dur (sec) & 1510   & 1744 & 2308 & 2296  \\ \hline
Mean $|d|$ of ABSs &  {0.0030}  & {0.0017}  & {0.0014}  & {0.0009}\\ \hline
Total dur (sec) & 5400 & 5400 & 7708 & 7708    \\ \hline
\end{tabular}
\end{center}
\end{table}

Table~\ref{tab:dsets} presents statistical characterization of the ground-truth ABS annotations on the arousal and valence contours of the RECOLA and CreativeIT datasets. Total ABS region durations cover around 30\% of the datasets with similar mean ABS durations of 3.6~seconds. Mean absolute delta ($|d|$) values for the ABS regions are observed higher for the RECOLA dataset that indicates higher affect contour changes for the RECOLA.

\subsection{Implementation Details and Setup}

Experimental evaluations of the proposed ABS detection system are executed using cross-validation. For the RECOLA dataset, 2 videos are selected as the test set and the rest are chosen for training and validation, resulting in 9 folds. On the other hand, we apply a leave-one-session-out strategy for the CreativeIT dataset, for each fold 1 session is chosen as test and rest are used for training and validation resulting in 5 folds.

We set the length, $L$, of the first order regression coefficients to temporally capture 0.8~seconds, the ABS temporal window size, $w_s$, is set to span 2~seconds. In (\ref{eq::annotation}), we set two threshold, $\tau$, values, one for each dataset, so to cover 30\% of the datasets as the ABS regions. 

The input of the KFDCNN, $I[n]$, is set with $T=100$ and $s=5$ which spans an 8~seconds temporal window with dilation rate $5$. Kernel fusion layer at KFDCNN has 3 parallel 1-dimensional convolutions with kernels sizes 3, 5, and 7, and these kernels have a dilation rate of $s=5$. The second 1-dimensional convolution has a kernel length of 3 with a dilation rate of 1. The max-pool layer down-samples the temporal dimension into half. The output of the max-pool layer is flattened from 2-dimension into 1-dimension and feed into fully connected layers with node sizes of 40, 20, and 2 respectively. 

Single kernel and no dilation derivatives of the KFDCNN are also defined and evaluated to better assess the performance of the proposed model. The CNN architecture, which is depicted within the red dashed lines in Figure~\ref{fig:KFDCNN}, has a single kernel set with size of 3 and dilation rate of 1. In order to have comparable complexity with the KFDCNN, input feature of the CNN is set with $T=20$ and $s=1$ which spans 1.6~seconds temporal window without dilation. A dilated CNN (DCNN) architecture is also defined by setting the input feature representation with $T=100$ and $s=5$.


\subsection{Affective Change Point Detection Performances}

Unweighted average F-score (UAF1) and Recall (UAR) metrics are computed at frame level via cross-validation and used for the performance evaluations. 

Table~\ref{tab:acd_f1} presents F1-score and Recall performances of the KFDCNN against the baseline FFN, CNN and DCNN. CNN-based architectures outperform the baseline in all comparison metrics by at least 2\% at RECOLA and CreativeIT databases. This result stresses the importance of temporal information for the detection of ABSs. Among the CNN based architectures, KFDCNN distinctly performs better than CNN and DCNN models. Performance improvement for the KFDCNN is highest for arousal and valence in the RECOLA dataset, and for valence in the CreativeIT dataset. 

Comparing CNN with DCNN, use of dilation improves the F-score performance by 5\% for arousal and 3\% for valence in the RECOLA dataset. Moreover, similar improvements are also seen with the CreativeIT dataset. Large temporal context due to dilated kernels is crucial to differentiate idle sections from ABS. This observation supports that consecutive temporal features carry less extrinsic information due to the slowly varying nature of emotional processes.

Comparing DCNN with KFDCNN, kernel fusion improves F-score approximately by 3\% for arousal and 2\% for valence at the RECOLA database. Similarly at the CreativeIT database, improvements are 2\% for valence. On the other hand, DCNN has only 0.2\% better performance for arousal at the CreativeIT database. This consistent improvement indicates that learning relationships at different temporal resolutions  improves ABS detection. 

\begin{table}[!htb]
\begin{center}
\caption{F-score and Recall performance results of the baseline (FFN), convolutional neural network (CNN), dilated convolutional neural network (DCNN), and the proposed KFDCNN framework for the affective burst detection over the RECOLA and CreativeIT datasets on Arousal (A) and Valence (V) contours}
\label{tab:acd_f1}
\resizebox{\columnwidth}{!}{
    \begin{tabular}{l|c c c c |c c c c}
    \toprule[1pt]\midrule[0.3pt]
    \textbf{Model} &\multicolumn{4}{c|}{\textbf{RECOLA}} & \multicolumn{4}{c}{\textbf{CreativeIT}}\\
    &\multicolumn{2}{c}{UAF1} &\multicolumn{2}{c|}{UAR}&\multicolumn{2}{c}{UAF1}&\multicolumn{2}{c}{UAR}\\
           & A       & V                     & A      &  V                          &  A     &  V    &  A &  V \\ \midrule \hline 
{FFN}     & {56.2}  & {53.7}                 & {56.8} & {54.4}                      & {48.1} & {51.0}                      & {55.8}           & {56.1}           \\ \hline 
{CNN}     & {59.0}  & {56.3}                 & {60.1} & {59.0}                      & {53.8} & {56.4}                     & {58.5}           & {59.6}           \\ \hline  
{DCNN}     & {64.3}  & {59.4}                 & {64.8} & {60.3}                      & {\textbf{57.2}} & {56.8}                     & {\textbf{60.3}}          & {58.0}           \\ \hline  
{KFDCNN} & {\textbf{67.0}} & {\textbf{61.4}}  & {\textbf{68.5}} & {\textbf{62.2}}    & {57.0} & {\textbf{58.5}}   & {59.4}  & {\textbf{60.0}}   \\ \hline 
\end{tabular}
}
\end{center}
\end{table}

KFDCNN has a superior performance on the RECOLA than CreativeIT, by approximately 10\% for arousal and by 3\% for valence. This could be due to the fact that RECOLA is not an acted dataset, and as a result it includes more spontaneous changes and exhibits higher mean absolute delta, $|d|$, values within ABSs compared to the CreativeIT. 

\section{Conclusion}

In this study, we present the affective burst detection as an important affective computing problem that can introduce the capability of capturing affective fluctuations better in the inter-and intra-emotion domain. We address the affective burst detection as an imbalanced binary problem of segmenting affective contour into burst and idle regions.

First, we label the affective contour by first detecting ABPs over the derivatives of the affective attributes. Later, the annotations are generated by extending the ABPs into segment vectors ABSs. The proposed KFDCNN architecture is trained with the generated annotation targets. From the conducted experiments, we observe that the KFDCNN outperforms the baseline architecture for F1-score and Recall on both RECOLA and CreativeIT datasets. Moreover, we depicted the importance of introduced concepts dilation and kernel fusion by comparing KFDCNN with CNN and DCNN. It is seen that larger receptive filed size due to dilation brings at least 3\% improvements, and kernel fusion brings at least 2\% improvements. Considering these observations, we suggest using KFDCNN for the affective burst detection.

\bibliographystyle{IEEEbib}
\bibliography{AffectiveChangePoint_Det}

\end{document}